Fast track to the overdoped regime of superconducting YBa2Cu3O7-δ thin films via electrochemical oxidation

Alexander Stangl[1,2,3*], Aiswarya Kethamkuzhi[4], Hervé Roussel[3], Cornelia Pop[4], Xavier Obradors[4], Teresa Puig[4], Mónica Burriel[3] and Arnaud Badel[5]

* alexander.stangl@grenoble-inp.fr

[1] Université Grenoble Alpes, CNRS, Grenoble INP, Institut Néel, 38000 Grenoble, France

[2] TU Wien, Atominstitut, Stadionallee 2, 1020 Vienna, Austria

[3] Université Grenoble Alpes, CNRS, Grenoble INP, LMGP, 38000 Grenoble, France

[4] Institut de Ciència de Materials de Barcelona (ICMAB-CSIC), 08193 Bellaterra, Barcelona, Spain

[5] Université Grenoble Alpes, CNRS, Grenoble INP, G2ELab – Institut Néel, 38000 Grenoble, France

**Abstract:**

High temperature superconductors, especially $YBa_2Cu_3O_{7-\delta}$ (YBCO), are considered a key enabling technology towards a clean energy future. Hole doping in YBCO is a prerequisite for the emergence of its unchallenged superconducting properties. Up to now, research was focused on the under- and optimally doped region, due to practical limitations in reaching the overdoped state, despite being highly interesting from fundamental and applied aspects as competing orders vanish and critical current densities are expected to peak. Here, we deploy for the first time an electrochemical method to access the mostly uncharted overdoped region. We demonstrate precise control over the bulk oxygen concentration in YBCO thin films across the full off-stoichiometry window ($0 \lesssim \delta \lesssim 1$) using electrochemical oxidation combined with *in situ* XRD and electrical measurements. Resulting high doping states and critical current densities are confirmed using a multi modal approach, including x-ray diffraction, electrical, Hall and magnetic characterization. Thus, this work opens a promising pathway based on electrochemical oxidation towards electronically clean, oxygen overdoped cuprate superconductors and therefore will assist to further push the critical current density to its intrinsic limit.

**1. Introduction**

Superconductivity is considered one of the key technologies towards a clean energy future, offering unique inherent advantages for various application scenarios, including all stages of modern energy infrastructure[1,2] (generation, transmission, grid resilience, storage and consumption), transport, medical applications and research. As an enabling technology, it opens the door to an entirely new and disruptive technological paradigm, such as compact fusion reactors,[3,4] to revolutionize the way we harvest energy.

Commercial implementation of high temperature superconductors (HTS) into applications requires the fabrication of so-called coated conductors (CC), based on the uniformly textured deposition of thick, nano-engineered layers in kilometres length, demanding control over more than 12 orders of magnitude: from the sub nm range to kilometres in length.[5–8] The resistance free transport current is given by the product of the material and layer quality dependent critical current density, $J_c$, and the cross section of the conductor. In practise, the superconducting layer thickness is limited to around 1-4 μm, due to the gradual loss of growth control. On the other hand, increasing $J_c$ is a promising approach to enhance performance, while simultaneously reducing the critical raw material consumption.[8,9] A large number of different defect landscapes has been engineered to introduce artificial pinning centres and optimize $J_c$ in distinct regions of the $(T, B)$ phase diagram.[8–15] Less attention has been drawn to maximise the superconducting condensation energy, $E_c$, itself, which sets the scale for the maximum achievable critical current density for a given defect landscape and is expected to peak in the so-called overdoped regime, as schematically shown in Figure 1(a) (for a definition of the term (over-)doping see SI-Note 1).[9,16–21]

In YBa$_2$Cu$_3$O$_{7-\delta}$ (YBCO), the most promising superconducting material for energy applications, the charge doping state can be controlled via chemical substitution of Y with aliovalent cations (such as Ca), or more commonly, via the variation of the oxygen off-stoichiometry, $\delta$. This oxygen deficiency can be modified by heat treatments under controlled oxygen partial pressure conditions in an easy manner, but with limited precision, flexibility and reproducibility due to the complex physicochemical processes involved.[22–24] Thus, research has primarily focused on the easier accessible underdoped and optimally doped region. While the introduction of job-sharing mechanisms via catalysts may provide kinetically faster reaction pathways,[25] thermodynamic limitations prevail (*e.g.* the $pO_2$ dependence of the oxygen concentration), and may be addressed via high pressure treatments.[26–29] Generally speaking, controlling, measuring and fine-tuning the oxygen concentration has been a long quest for many branches in physics and materials science, as it dominates the relationship between defects, structure and functional properties in many different oxides. The recent renaissance of electrochemical titration processes may ultimately tackle these issues by providing a revolutionary way to (i) outsource limiting elementary surface reactions from the oxide of interest (here YBCO) to other functional materials, which outperform these tasks, (ii) shift the thermodynamic oxygen defect state to levels far beyond reach for conventional experimental setups via the application of an electrochemical overpotential and (iii) quantify changes in the oxygen stoichiometry via coulometric titration. This approach implies a paradigm shift for the precise control of ionic defect densities in strongly correlated, functional oxides.[30] It allows to study and tune fundamental electrical, optical and magnetic properties with unprecedented flexibility and precision,[31,32] including metal-insulator and topotactic phase transitions,[33,34] and is fundamental to novel applications, such as oxygen storage technology,[35,36] oxygen ion batteries,[37,38] synaptic memory devices,[39] and "Motttronics".[40]

In contrast to recently emerging field-effect based experiments,[41–43] where the oxygen vacancy landscape is locally rearranged in narrow gate areas by high electric fields (you may refer to this as *brute force* approach), the electrochemical method[*] discussed here holds several key advantages, including precise and measurable control over changes in the oxygen stoichiometry across the full sample volume using comparably small electric field strengths.

Notably, similar electrochemical methods have already been deployed in the very beginning of the HTS era to drive the superconducting transition in YBCO and $Bi_2Sr_2Ca_1Cu_2O_{8+x}$ (BSCCO) bulk, as well as thin films.[44–46] However, the lack of high quality epitaxial superconducting material at that time prevented full exploitation of its capacity to investigate the doping dependence of $J_c$ and go beyond the optimal doping state. Thus, it is yet for us to demonstrate its far-reaching potential for the field of superconductivity by bridging the gap towards the overdoped regime, where competing orders vanish and superconducting critical current densities are expected to peak.

Here, we use for the first time electrochemical oxidation to access the mostly uncharted overdoped region of high quality, epitaxial $YBa_2Cu_3O_{7-\delta}$ superconducting thin films and verify its capability to flexibly tune the oxygen stoichiometry (across the full oxygen doping window ($0 \lesssim \delta \lesssim 1$) as well as the full sample volume) and achieve high critical current densities based on a multi-modal approach, including *in situ* XRD, combined electrochemical and electrical measurements as well as low temperature characterisation techniques. Thereby this study lays the groundwork for future understanding of the doping dependence of the critical current for a given defect structure and the exploration of new physics in electronically clean overdoped YBCO.

---

[*] An electrochemical mechanism involves ionic charge transfer across interfaces, such as the electrolyte and electrodes, while electrostatic ones are limited to rearrangements of ions within the material, such as during the formation of an electric double layer, or local changes in oxygen stoichiometry due to field-enhanced vacancy mobility.

## 2. Methods

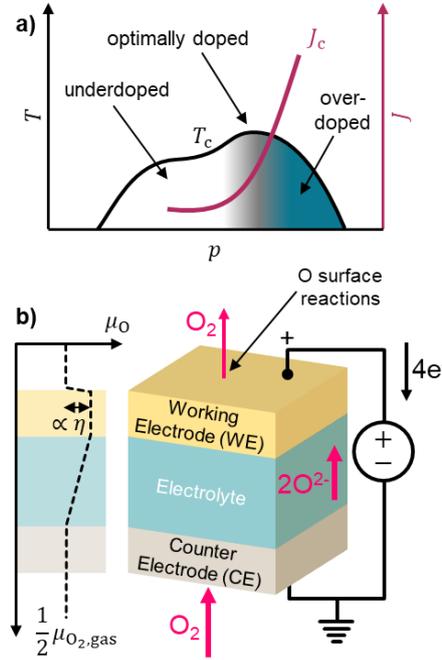

Figure 1: (a) Cuprate phase diagram for critical temperature, $T_c$ and critical current density, $J_c$, schematically reproduced from literature.[20,21] (b) Principle of electrochemical oxidation of the working electrode (WE) in an electrochemical cell and schematic representation of the oxygen chemical potential, $\mu_O$ profile across the cell (with $\mu_O = 0.5\mu_{O_2}$ for an ideal gas). Note, that $\mu_O$ must drop across the WE surface for the successful application of an overpotential, which is achieved if a WE surface reaction is limiting the overall oxygen flux across the cell. The width of the pink arrows schematically represents the magnitude of the flux.

In oxide materials with flexible oxygen stoichiometry, the equilibrium oxygen content is determined by the chemical potential of oxygen, $\mu_O$, which depends on they oxygen potential of its surrounding, $\mu_{O,\text{ref}}$ (via the $pO_2$ of the atmosphere and the temperature) and any applied electrochemical overpotential, $\eta$:[47]

$$\mu_O(T) = \mu_{O,\text{ref}}(T) + 2e\eta \qquad \text{Eq. 1}$$

Conventionally, only $\mu_{O,\text{ref}}$ is modified to vary the oxygen content, with practical limitations in terms of experimental setups and procedures (e.g. $pO_2 \leq 1\text{atm}$, finite annealing times to reach equilibrium, *etc*). The application of an overpotential, however, vastly extends the accessible oxygen off-stoichiometry range and allows precise tuning of the oxygen content. In mixed ionic electronic conducting (MIEC) thin film systems, oxygen surface reactions are commonly found to limit the overall oxygen exchange process,[22] which makes them act like a chemical capacitor. Thus, applying a voltage across an electrochemical (EC) cell, consisting of an ionic conducting electrolyte, sandwiched between two active electrodes, as shown in Figure 1(b), allows to establish an overpotential in the MIEC working electrode (WE) and change its oxygen stoichiometry.[30,45] The application of a positive voltage drives an electrochemical current from the counter electrode (CE) towards the working electrode, surpassing a cascade of transport and exchange processes: oxygen is incorporated from the atmosphere at the CE and diffuses as ion in the form of $O^{2-}$ through the electrolyte and across the interface into the MIEC thin film.[48] While some of the ions remain in the crystal structure of the WE, effectively increasing its oxygen stoichiometry and building up the voltage of the chemical capacitor, the rest is removed through the WE surface, corresponding to a leakage current in the steady state, $I_{\text{leak}}$. Applying a negative bias results in an electrochemical current in reverse direction, whereas the oxygen concentration of the MIEC is reduced. These thermally activated processes are sustained only at sufficiently high temperatures ($\gtrsim$ 250°C). Charge neutrality in both electrodes is maintained via an electronic current over the external circuit, which allows to estimate the change of the oxygen off-stoichiometry, $\Delta\delta$, of the WE via coulometric titration:[49]

$$\Delta\delta = \frac{V_{\text{uc}}}{2e \cdot V_{\text{film}}} \int (I(t) - I_{\text{leak}})dt \qquad \text{Eq. 1}$$

with the unit cell and thin film volume, $V_{\text{uc}}$ and $V_{\text{film}}$, respectively. The oxygen activity, *i.e.,* the effective oxygen partial pressure inside the material, assuming ideal gas behaviour, is given via the Nernst equation:[50]

$$pO_{2\,\text{eff}} = pO_{2\,\text{ref}} e^{\frac{4F\eta}{RT}}, \qquad \text{Eq. 2}$$

with the oxygen pressure inside the electrochemical setup, $pO_{2\,\text{ref}}$, and the Faraday and gas constants, $F$ and $R$. The overpotential can be determined via:[51]

$$\eta = U_{\text{appl}} - I_{\text{leak}}R, \qquad \text{Eq. 3}$$

with the polarization resistance, $R$, of all cell components but the working electrode. If CE surface reactions and electrolyte ion conduction are comparably fast, the applied voltage almost entirely translates into an overpotential and one can approximate $\eta \approx U_{\text{appl}}$.

This simple method holds several key advantages. The full bulk oxygen defect landscape can be controlled with high flexibility and precision using a standard voltage source and the experimentally accessible oxygen off-stoichiometry range is vastly extended.[30] Sluggish surface reactions can be outsourced to better performing functional materials, which allows to strongly shorten annealing times and minimize associated materials degradation and ageing processes. High effective oxygen partial pressures only affect the oxygen sub-lattice, without the mechanical impact of high pressure treatments on the crystal structure and compression induced deformation of ionic bonds and antagonistic effects such as reduced stability of high oxidation states and pressure decreased diffusion rates.[26,52] Furthermore, electrochemical methods allow to project well-defined oxygen gradients into a single thin film, with great potential for fundamental studies of the emergence of the superconducting phase.[53]

## 3. Results and discussion

In this work, we studied dense, 200 nm and 250 nm thick sub-stoichiometric YBa$_2$Cu$_3$O$_{7-\delta}$ thin films, with $0 \leq \delta \leq 1$. The samples were synthesized either by chemical solution deposition (CSD) or pulsed laser deposition (PLD) on top of 5×5 mm ionic conducting yttria-stabilized zirconia (YSZ) single crystal substrates (100), serving as electrolyte. The device architecture is depicted in Figure 2(d). An epitaxial $Ce_xZr_{1-x}O_{2-\delta}$ (CZO) buffer layer, grown by CSD,[54] enabled the heteroepitaxial (00l) growth of YBCO, as confirmed by X-ray diffraction (XRD), see Figure 2(a) and SI-Figure 1(a) for PLD and CSD grown YBCO, respectively. For the case of CSD, the presence of the $BaCeO_3$ phase is a sign of persisting reactivity between YBCO and the CZO buffer layer, which however did not impede the epitaxial growth. SEM top surface images confirm the flat, homogenous structure of the multilayer systems showing the typical surface morphology obtained from those methods, *cf.* Figure 2(b & c) and SI-Figure 1(b & c). A porous, paint brushed Ag layer served as counter electrode at the bottom of the electrolyte supported cell. For good electrical current collection and homogeneous polarization, an Au layer was brush painted on top of YBCO.

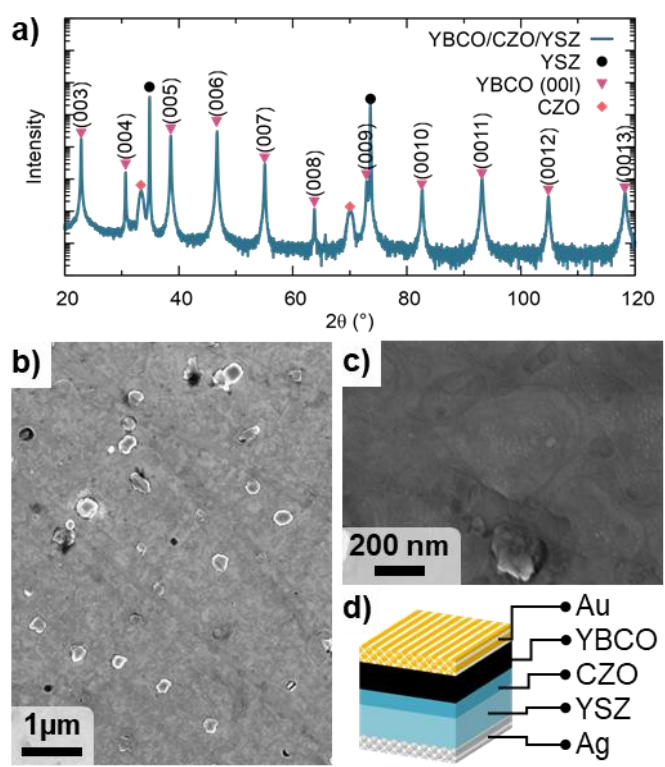

Figure 2: (a) XRD diffraction pattern confirming highly epitaxial growth of YBCO (PLD) on YSZ substrate using $(Ce,Zr)O_{2-\delta}$ buffer layer. (b & c) Secondary electron SEM images of YBCO (PLD) surface. The architecture of the electrochemical device with a porous Au current collector on top of YBCO and a porous Ag counter electrode is shown in (d).

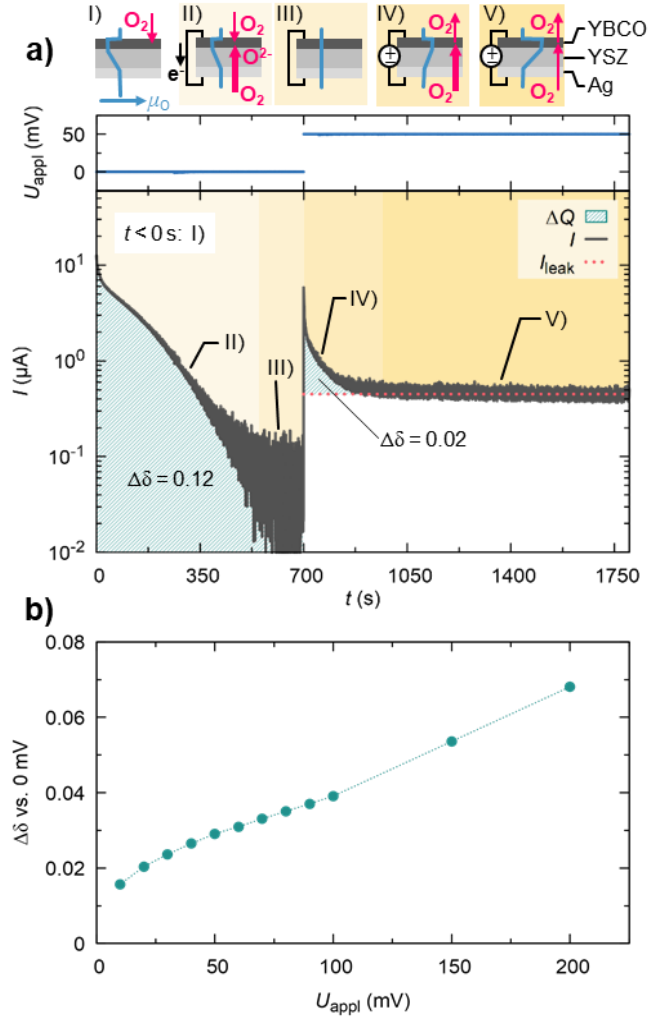

Figure 3: (a) Outsourced oxygen incorporation and electrochemical oxygen pumping into as-deposited YBCO (PLD) at 380 °C in 1 atm $O_2$. Main and top panel show the measured current and the applied voltage across the electrochemical cell, respectively. Here, oxygen incorporation takes place in five stages (see also schematics on top): I) via YBCO surface reactions ($t < 0\,s$) and II) through the Ag counter electrode upon short-circuiting the electrochemical cell until III) reaching the thermodynamic equilibrium with the atmosphere. IV) Application of an overpotential (+50 mV) drives additional oxygen into the YBCO layer beyond the chemical equilibrium (defined by $pO_2$ and $T$), quantifiable via the marked area under the $I(t)$ curve, until V) electrochemical saturation is reached with a constant oxygen flux through the cell in the steady state. The thickness of the pink arrows in the schematics on top represent the magnitude of the net oxygen flux, the blue lines illustrate the oxygen chemical potential profiles during the different phases. (b) Change of the oxygen off-stoichiometry of CSD grown YBCO film as function of applied bias, relative to the unbiased state at 375 °C in 1 atm of flowing $O_2$.

**Electrochemical oxygen control**

An electrochemical oxidation process is depicted in Figure 3(a) for an as-grown YBCO sample (PLD). One can distinguish five distinct phases (I – V), illustrated on top of Figure 3(a) and discussed in the following. YBCO is grown under strongly reducing conditions, resulting in high oxygen deficiencies of as-grown samples with typical off-stoichiometries of about $\delta \approx 0.5 - 0.6$. A following annealing in oxygen rich atmosphere is therefore characterised by a strong chemical driving force for oxygen incorporation due to a large step in oxygen chemical potential between the sample and the atmosphere (as indicated using the blue line in the schematics on top of Figure 3(a)). This initial stage of oxygen incorporation (phase I) takes place via a sluggish gas-solid reaction, namely the oxygen reduction reaction (ORR) at the native YBCO surface, and corresponds to the standard oxygenation process. It is generally limited by the experimentally accessible $pO_2$ conditions, as well as slow surface kinetics of YBCO. This phase can be studied using conventional electrical measurements,[25] but was not monitored here.

Upon electrically short-circuiting the YBCO layer with the porous Ag counter electrode at $t = 0$ (phase II), oxygen surface reactions are *outsourced* to the kinetically more active CE, followed by fast diffusion of oxygen ions, $O^{2-}$, through the YSZ electrolyte and across the interface into the YBCO layer, as this reaction pathway is overall faster than mixed ionic electronic transfer reactions at the YBCO surface. Charge neutrality in both electrodes is maintained via the measureable backflow of electrons through the external circuit, corresponding to the shown $I(t)$ curve in the main panel of Figure 3(a). The area under the curve is proportional to the change in oxygen stoichiometry of YBCO, $\Delta\delta = 0.12$. Note that this number only accounts the oxygen transported through the electrolyte. Simultaneously, oxygen continues to enter through the native YBCO surface. As this process however is considerably slower, its contribution to the total change in oxygen stoichiometry of the YBCO layer is neglectable. Once YBCO is in chemical equilibrium with the surroundings, defined by a homogenous $\mu_O$ across the cell and the atmosphere, the net oxygen flow vanishes and accordingly the current decays to zero (phase III).

The application of a voltage (+50 mV, phase IV) shifts the chemical potential inside the MIEC out of its balance with the gas phase and reinitiates a net oxygen flow, characterised by the sharp rise in measured current. The coloured area under the curve shows that additional oxygen ($\Delta\delta = 0.02$) is incorporated into the YBCO layer, exceeding the chemical equilibrium state at given $T$ and $pO_2$ conditions. The current decays towards a constant level in the steady state (phase V), as the stoichiometry of the MIEC approaches electrochemical saturation. This leakage current, $I_{\text{leak}}$, is a measure of the net oxygen flux limited by the RDS of the YBCO surface (under given $\eta, T, pO_2$ conditions) and corresponds to the continuous pumping of oxygen through the electrochemical cell (see schematic on top). It arises due the step in $\mu_O$ at the interface between the atmosphere and YBCO, as the effective $pO_2$ inside the WE is higher than the real $pO_2$ in the gas phase. $I_{\text{leak}}$ is about 400 nA, and therefore about 2 orders of magnitude smaller, than the maximum incorporation rate at the counter electrode (≈ 10 µA, *cf.* at $t = 0$ and 700 s).

Increasing the applied voltage (and thus the oxygen chemical potential of YBCO) allows to gradually increase the oxygen concentration, as shown in Figure 3(b). The corresponding $I(t)$ curves are given in SI-Figure 2. The rapidly raising $I(U)$ dependence above 100 mV, displayed in the inset of SI-Figure 2, may reduce the efficiency of translating the voltage into an overpotential and in turn, could potentially limit the maximum achievable oxygen stoichiometry, as discussed in more detail in the next section.

Continuous high oxygen fluxes (corresponding to the measured high currents) through the YBCO structure may additionally impact the sensitive superconducting properties and therefore should be minimised. Different strategies to counter such high leakage currents are currently being investigated.

**Flexible tuning of the bulk oxygen stoichiometry**

At this point, the presented results already highlight two major advances of the chosen method: first, sluggish surface kinetics of YBCO are bypassed and oxygen gas incorporation is outsourced to a better performing counter electrode. Second, utilizing an overpotential allows to increase the oxygen stoichiometry of YBCO beyond the chemical equilibrium defined by temperature and $pO_2$. In the following we combine electrochemical oxidation with simultaneously performed kinetic *in situ* X-ray diffraction measurements to verify its capability to flexibly tune the oxygen concentration across the full volume of the bulk. We take advantage of the fact that the unit cell lattice parameters of various oxides are highly sensitive to the oxygen stoichiometry.[55–57] This can be understood by a combination of an increase of the ionic radius upon reduction of the cation as well as electrostatic repulsion of the positively charged vacancies from the positive cation sub-lattice surrounding it.[58–60] We tracked the chemical expansion by monitoring the evolution of the (005), (006), (0012) and (0013) peaks at approximately 380 °C in $O_2$ atmosphere, while simultaneously ramping an applied voltage stepwise from -1 V to +1 V with steps of 100 mV every 30 min. The spectra acquisition time for both selected 2θ ranges of 33 – 50 ° and 100 – 120 ° was about 160 s each.

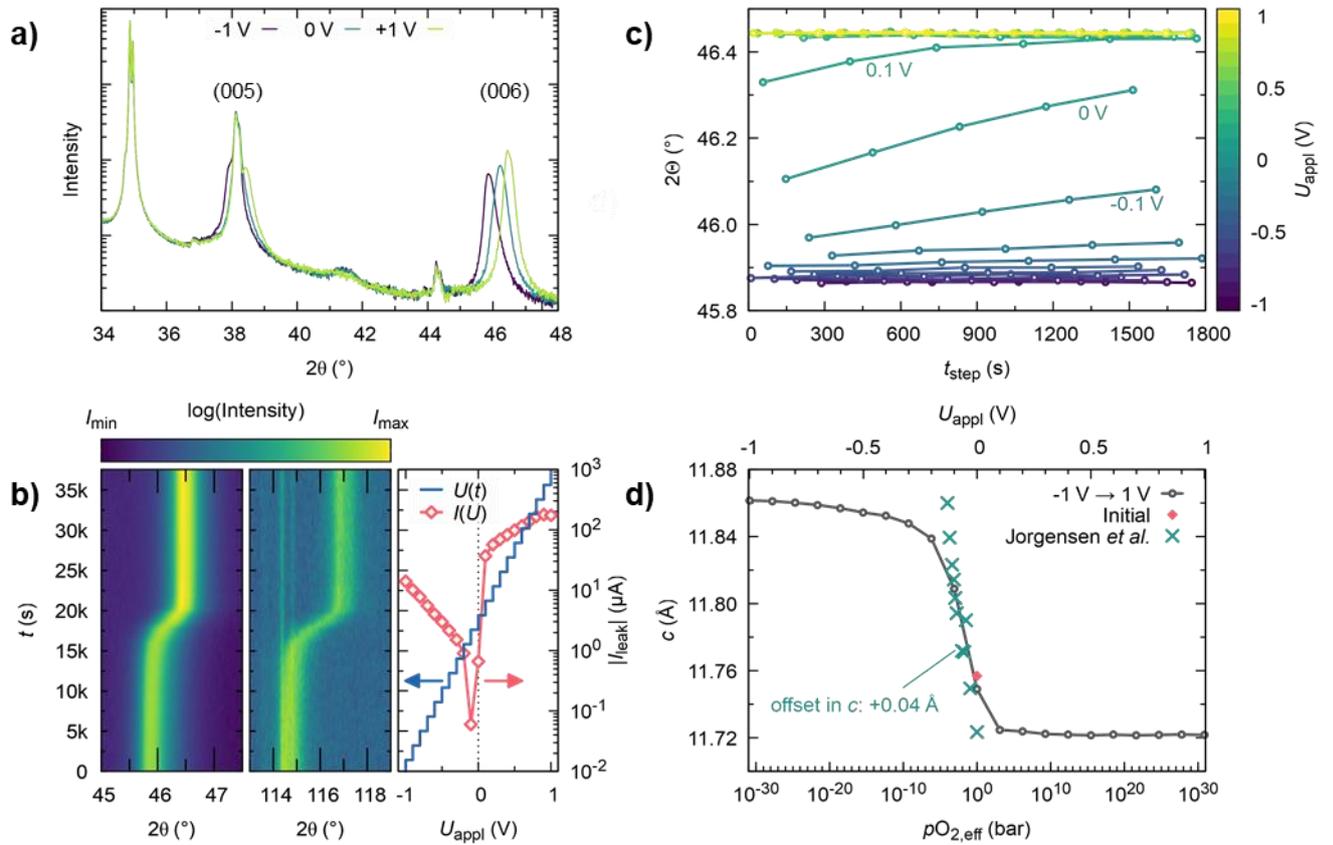

Figure 4: *In situ* XRD analysis of YBCO (CSD) during electrochemical pumping of oxygen at 380 °C in flowing oxygen: (a) End (-1, +1 V) and middle point (0 V) diffraction pattern for substrate and (005) and (006) YBCO peak. (b) Time resolved intensity maps of (006) and (0013) peak in the left and middle panel, respectively, under stepwise increase of the applied polarization from -1 to +1 V. The $U(t)$ curve is shown in the right panel (blue line), which also depicts the voltage dependence of the leakage current (pink curve). (c) (006) peak shift for each voltage step. (d) YBCO c-axis parameter as function of the effective oxygen partial pressure, compared to room temperature literature data,[56] (shifted upwards by +0.04 Å to compensate for thermal expansion). The pink data point represents the value obtained at 380 °C prior to any polarization.

Diffraction pattern of the endpoint and midpoint, corresponding to -1, 0 and +1 V, are shown in Figure 4(a) and the corresponding electrochemical titration curve is displayed in SI-Figure 4(c). Stable measurement conditions (*e.g.* temperature and sample alignment) are confirmed by the fact that the YSZ substrate (2Θ ≈ 35 °) and the $BaCeO_3$ impurity phase (peak at 44.3 °) remain unaffected by the bias, while YBCO (00l) reflections move to higher angles. We find tremendous shifts in 2Θ for the YBCO (00l) thin film peaks with bias, ranging from about 0.5 ° for the low angle reflections to more than 2 ° for the (0013) peak, as detailed in the time resolved intensity maps in Figure 4(b). The voltage profile, $U(t)$, is plotted in the right panel of Figure 4(b) (blue line). The full *in situ* data set can be found in SI-Figure 5. Notably, the YBCO (00l) peak widths remain approximately constant, which indicates that the oxygen concentration within the probed, macroscopic thin film volume is homogeneous in both, lateral and vertical direction (spot size of about 1 mm).

The position of the maximum of the (00l) peaks was determined via fitting using Voigt-profiles, see SI-Figure 4(a). The obtained time evolution of the (006) reflection is shown in SI-Figure 4(b) and as a function of the step time for each voltage step in Figure 4(c). From -1 to -0.4 V, only minor changes in peak position are observed, indicating that the material is close to the fully reduced state, *i.e.* $YBa_2Cu_3O_6$, with highly depleted oxygen chain sites. Such cathodic polarizations cause oxygen to be incorporated at the YBCO surface level and pumped through the electrolyte to the CE, with small net flow rates (cf. $I(U)$ curve, depicted with red diamonds in the right panel of Figure 4(b)). With increasing bias (*i.e.* less reducing condition), we find small but gradual changes in the diffraction pattern due to the incorporation of bulk oxygen into the YBCO crystal structure. The most prominent peak shifts occur in the narrow polarization range from -0.1 to +0.1 V, followed by saturation above 0.2 V. This saturation may arise either due to a structural limit, *i.e.* all possible oxygen sites are filled, or could be a consequence of the very high leakage currents of up to 200 µA under strong anodic biases (see the right panel of Figure 4(b)). In the latter case, the increase in voltage may be entirely consumed by a higher oxygen flux through the surface and therefore does not increase $\mu_O$ any further, which results in a stagnation of the oxygen stoichiometry. Electrochemical impedance spectroscopy (EIS) reveals a strongly reducing polarization contribution of YBCO with increasing bias, see SI-Figure 3, which indeed hampers the unlimited increase of the overpotential as oxygen diffusion through the YSZ and/or the oxygen reduction at the Ag counter electrode are gradually becoming co-limiting.

The high temperature *c*-axis parameter can be obtained from the measured 2Θ position using the Bragg equation and the dependence of the *c*-axis parameter on the effective oxygen partial pressure is plotted in Figure 4(d), whereas the correlation between $pO_{2,\text{eff}}$ and the applied overpotential is given by the Nernst equation (under the assumption $\eta \approx U_{\text{appl}}$, which may be violated for high biases). It is important to note, that the hugely varying $pO_{2,\text{eff}}$ (from $10^{-31}$ to $10^{31}$ bar) is not a physical pressure on the sample but solely the equivalent pressure corresponding to the oxygen chemical potential felt by the MIEC. Structural deformation and other detrimental effects due to compression are therefore avoided, while still being able to set the associated oxygen contents. At 380 °C, the lattice parameter varies continuously by more than 1 % from $c = 11.86$ Å at $10^{-31}$ bar (-1 V) to 11.72 Å at $10^{31}$ bar (+1 V) with a large absolute change of $\Delta c = 0.14$ Å.

As the linear thermal expansion coefficient in YBCO thin films depends only weakly on the oxygen off-stoichiometry,[61] $\Delta c$ is basically temperature invariant, which allows for comparison with literature. Room temperature data of the bulk YBCO *c*-axis reported by Jorgensen *et al* was added to Figure 4(d), using a constant offset to account for the thermal expansion of the unit cell. They found the same $\Delta c$ range at RT (and *in situ* at elevated temperatures) for samples with oxygen stoichiometries close to $O_6$ and $O_7$, as measured by iodometric titration and thermogravimetric analysis (TGA).[56,62] This suggests that the bulk off-stoichiometry, $\delta$, of our sample was equally modulated within a large part of the thermodynamic stability window from 0 to 1. The steeper $pO_2$ dependence of the YBCO bulk literature data is possibly a consequence of the higher annealing temperature (520 °C versus 380 °C), as well as the different sample type, *e.g.* thin films are known to be highly defected and can follow a modified oxygen phase diagram.

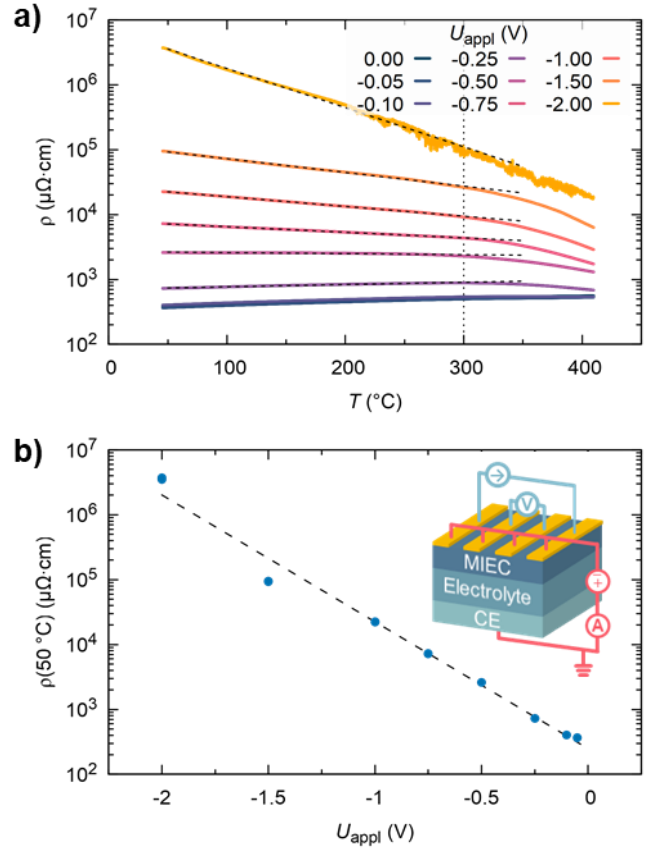

Figure 5: (a) Temperature dependence of the electrical resistivity of CSD grown YBCO thin film upon heating for different oxygen stoichiometries, which were set during prior electrochemical treatments at 410 °C in 1 bar of $O_2$ using polarizations from -50 mV to -2 V. (b) Electrical resistivity at 50 °C as function of the applied voltage (at 410 °C). The dashed line corresponds to an exponential dependence. The two electrode configurations are schematically drawn in the inset.

To demonstrate the intricate relationship between the oxygen stoichiometry and the electronic structure, we studied the high temperature electrical resistivity after application of different biases. Therefore, we deployed a special electrode geometry, using four metallic top electrodes (see inset in Figure 5(b)), which can be short-circuited for the out-of-plane polarization process, or individually contacted for in-plane resistivity measurements using a digital relay.[44,63] For details see SI-Figure 6(a & b). This allows to electrochemically set different oxygen stochiometries at high temperature (*i.e.* where ionic diffusion through the electrolyte is sufficiently fast, *e.g.* for YSZ ≳ 250 °C) and follow the resulting changes in functional electrical properties at lower temperatures within the same setup and without further sample handling. The programmed cycling of temperature and electronic configuration is illustrated in SI-Figure 6(b). Note that in the following we remove oxygen from the YBCO layer, as the transition from the optimally to the underdoped region is accompanied by drastic electronic changes already observable at high temperatures, rendering sample transfers between high and low temperature setups unnecessary.

A sample was prepared with the aforementioned top electrode configuration and was cathodically polarized at 410 °C until the current reached steady state. The bias was maintained during the following rapid cooling (100 °C·min$^{-1}$) until the ionic out-of-plane current vanished (around 250 °C), see SI-Figure 6(c). Then, the electrode configuration was automatically switched to measure the electrical resistivity of the thin film, while the temperature was further decreased down to 50 °C (with 50 °C·min$^{-1}$). Subsequently, the sample is slowly heated (20 °C·min$^{-1}$) back to 410 °C while monitoring the temperature dependence of $\rho$. Upon reaching 410 °C, the electrode configuration is switched back to the polarization one and the next bias is applied. An exemplarily resistivity measurement of a cooling-heating cycle is presented in SI-Figure 6(d) following a polarization of -100 mV.

The $\rho(T)$ heating curves following several different cathodic voltages are summarized in Figure 5(a). Close to the optimally doped state, YBCO exhibits metallic-like behaviour with a linear, positive temperature dependence, as observed for small biases. With increasing magnitude of the preceding negative polarization, and thus reduced oxygen concentration, the resistivity is increased at all temperatures and $\rho(T)$ curves approach exponential behaviour (black dashed lines), which is accompanied by a metal-insulator transition between -0.25 V and -0.5 V, indicated by a change of sign of the slope. While there is a strong chemical driving force for oxygen incorporation in this highly oxygen deficient sample, sluggish surface activity prevents sufficient oxygen influx at low temperatures, thus allowing accurate resistivity measurements (confirmed by the superposition of heating and cooling curves, *cf*. SI-Figure 6(d)). However, thermal activation of surface reactions causes the observed downward bending of the resistivity at around 300 °C, which marks the onset temperature for oxygen incorporation to equilibrate with the oxygen rich atmosphere, as described recently elsewhere.[25] Going from 0 V to -2 V, the electrical resistivity at 50 °C is increased over four orders of magnitude, due to the loss of charge carriers in the structure, as shown in Figure 5(b). Based on the deployed four-point measurement technique, these values correspond to the full bulk of the sample and verify that functional properties of the YBCO thin film can be flexibly tuned within an exceptionally wide range.

**Reaching the overdoped regime**

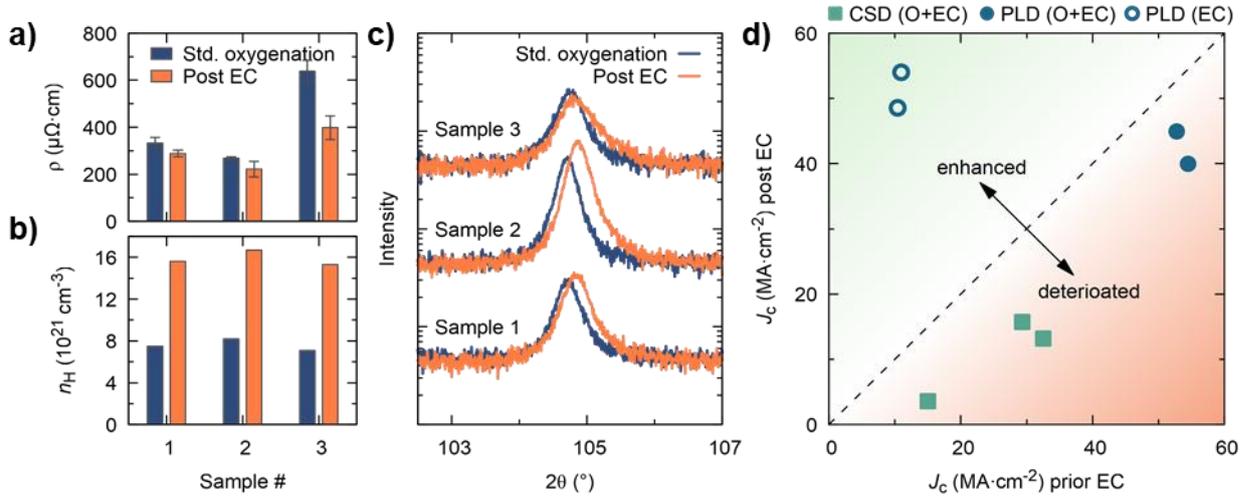

Figure 6: Change of (a) electrical resistivity and (b) charge carrier density and (c) shift of the (0012) XRD peak of YBCO upon electrochemical (EC) treatment for prior optimally doped YBCO (CSD) (all values correspond to room temperature). (d) Comparison of the critical current density at 5 K prior (*x*-axis) and post (*y*-axis) electrochemical treatment for YBCO grown by CSD (squares) and PLD (circles). Samples represented by filled symbols were already optimally doped via a standard oxygenation before the EC process (O+EC), open symbols indicate that the EC oxidation was the first (and only) oxygen treatment.

Oxygen overdoping is considered a promising complementary path towards outperforming superconducting properties.[21] Furthermore, it allows to suppress competing orders, such as the pseudogap, and investigate the emergence of superconductivity in electronically clean samples.[64,65] Yet, this region of the phase diagram was seldom accessed by purely anionic doping and approaches such as Ca-substitution and high pressure treatments are linked to detrimental effects on the structure and critical currents.[17]

In the following, we analyse the capability of electrochemical oxidation to access the overdoped state and its influence on electrical and superconducting properties. As a direct measure of the electronic doping (*i.e.* the holes per Cu ion in the Cu-$O_2$ planes) is challenging, we use the overall charge carrier density as a descriptor of the doping state, whereas optimal doping corresponds to about $n_H(T = 300K) \approx 8 \cdot 10^{21}$ cm$^{-3}$ (see SI-Note 1).[16,66,67] First, we investigated CSD grown thin films, which were formerly oxygenated using a conventional oxygenation process.[68,69] Starting from an optimally doped state, a subsequent electrochemical oxidation successfully decreased the electrical resistivity and doubled the number of free charge carriers in all studied samples, see Figure 6(a & b), reaching unprecedented high values of $n_H \approx 16 \cdot 10^{21}$ cm$^{-3}$ in CSD-grown YBCO thin films. We further observed a systematic shift of the (00l) peaks to higher angles, as exemplarily shown in Figure 6(c) for the (0012) reflection and in SI-Figure 7 for the full spectra, indicating a shrinkage of the c-axis lattice parameter. Experimental details and corresponding numerical results are given in Table 1. These findings indicate that we systematically increased the oxygen concentration and thus the electronic doping in all samples. Interestingly, the critical temperature only slightly decreased by ≤1 K, *cf.* Table 1, as was observed previously in other cuprate systems.[26]

| Sample | $T$ | $U_{appl}$ | $\rho$ (μOhm cm) | | $c$ (Å) | | $n_H$ ($10^{21}$ cm$^{-3}$) | | $T_c$ (K) | |
|---|---|---|---|---|---|---|---|---|---|---|
| | °C | mV | initial | final | initial | final | initial | final | initial | final |
| 1 | 360 | 700 | 333.3 | 288.5 | 11.674 | 11.666 | 7.5 | 15.6 | 90.7 | 90.4 |
| 2 | 360 | 200 | 268.2 | 221.4 | 11.675 | 11.662 | 8.2 | 16.7 | 91 | 89.9 |
| 3 | 400 | 400 | 638.4 | 398 | 11.671 | 11.669 | 7.1 | 15.3 | 89.7 | 88.9 |

Table 1: Overview of experimental parameters (temperature, *T*, and applied voltage) of the electrochemical oxidation and characterization results, incl. room temperature resistivity, c-axis parameter and charge carrier density, as well as the critical temperature before and after the electrochemical treatment in O$_2$ atmosphere.

The critical current density, prior and post electrochemical oxidation, is compared in Figure 6(d) (green filled squares) and tabled in SI-Table 1. Despite an enhancement in the doping state, the critical current densities were deteriorated for previously oxygenated samples. A similar degradation was found as well for previously oxygenated PLD-grown samples upon application of an overpotential, see filled circles in Figure 6(d). However, this effect may not be directly related to the electrochemical process: previously we observed that the simple re-exposure of the material to standard oxygen annealing conditions resulted in reduced critical currents, as detailed in SI-Figure 8,[67] which may be related to the known susceptibility of YBCO to water and other ambient contaminants.[70]

To separate the effects of electrochemical oxidation and re-annealing in oxygen, we prepared a batch of PLD samples, with no prior oxygen treatment, *i.e.* cooled directly after the synthesis in low $pO_2$ growth atmosphere (as-deposited). The resulting samples were highly oxygen deficient, with a low critical temperature (around 55 K) and high electrical resistivity, *cf.* Figure 7(a). The following first anneal in oxygen rich environment was assisted by electrochemical overpotentials to increase the oxygen stoichiometry beyond the equilibrium according to 1 bar of O$_2$ in the atmosphere (as shown in Figure 3(a)). The successful incorporation of oxygen into the thin film is confirmed by the strong increase of $T_c$ to about 89.5 K and a drop in $\rho(T)$, as shown in Figure 7(a). Hall measurements revealed a fourfold increase of the charge carrier density at 100 and 300 K, see the inset in Figure 7(b), reaching $n_H(100\,\text{K}) \approx 6.5 \cdot 10^{21}$ cm$^{-3}$ and $n_H(300\,\text{K}) \approx 15 \cdot 10^{21}$ cm$^{-3}$, respectively. The successful enhancement of the critical current density is clearly seen in Figure 7(b), with high absolute $J_c$ values at 5 K and 77 K of up to 54 MA·cm$^{-2}$ and 2.7 MA·cm$^{-2}$, respectively. The similar critical current densities found for the initial state of the standard oxygenated PLD sample (prior EC) and the final state (post EC) of the directly electrochemically treated sample with increased oxygen content (compare open and filled circles in Figure 6(d) and values given in SI-Table 1 for sample 4 and 6) calls for a more systematic study on the doping dependence of the critical current, which is currently being undertaken. Nevertheless, our results demonstrate for the first time that the chosen electrochemical oxidation procedure is compatible with obtaining high $J_c$ values in YBCO, while providing a reliable path towards overdoping of the crystal structure, rarely accessible by other means. Therefore, this method opens the possibility to maximise the achievable superconducting current for a given microstructure and pinning defect landscape.

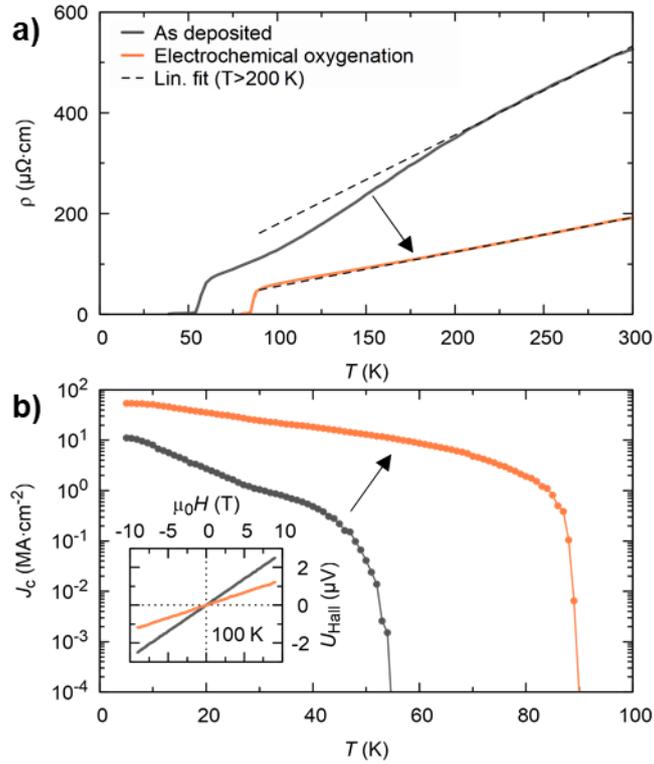

Figure 7: Low temperature physical properties of as-deposited and electrochemically oxygenated PLD YBCO thin film: (a) electrical resistivity measured in Van-der-Pauw configuration and (b) critical current density obtained via SQUID magnetometry. Inset in (b) shows the Hall voltage measured between -9 and +9 T at 100 K.

# Conclusions

In conclusion, we proposed a promising electrochemical method to lift experimental thermodynamic and kinetic restrictions for the optimization of the oxygen defect landscape of cuprate superconductors with unprecedented precision and flexibility. Electrolyte supported electrochemical cells were fabricated with epitaxial YBCO superconducting thin films as working electrodes using the PLD and CSD growth technique. The oxygen stoichiometry was tuned via the application of an electrical voltage across the cell and the doping dependence of structural and functional properties was investigated using *in situ* XRD and electrical measurements. Deploying electrochemical oxidation, we reached very high doping states in YBCO thin films as confirmed by Hall effect measurements and high critical current densities, marking a possible turning point for the development of out-performing superconductors. This method will allow for an unparalleled study of the doping dependence of normal state and superconducting properties and heralds a new era for the oxygen defect control in superconductors to enable the exploration of new physics and possibly enhance superconducting materials properties, highly relevant for both, basic and applied research.

# Experimental

All thin films studied within this work were synthesized at ICMAB using the chemical solution deposition (CSD) technique,[71,72] as well as pulsed laser deposition (PLD) on 5×5 mm² yttrium stabilized zirconia (001) (YSZ) single crystal substrates (Crystec) with an epitaxial $Ce_xZr_{1-x}O_{2-\delta}$ (CZO) buffer layer grown as well by CSD.[54] For the CSD growth of YBCO, anhydrous TFA precursor salts were dissolved in trifluoroacetic acid, trifluoroacetic anhydride and acetone to obtain a 0.25 mol (with respect to Y) YBCO-TFA solution, which was deposited by spin-coating on the substrate. The organic content was removed in a pyrolysis step at 310 °C for 30 min in humid, oxidising atmosphere using slow heating and cooling ramps between 3 and 5 °C·min⁻¹ to avoid the formation of cracks. The subsequent growth and sintering process was performed at 810 °C in 200 ppm of $O_2$ in flowing $N_2$ gas. For the nucleation stage, the gas is humidified to enable the trifluoroacetate based chemical growth of YBCO, whereas sintering is performed in dry atmosphere. A detailed description of the chemical reactions taking placed during the pyrolysis and growth process can be found elsewhere.[68,73] PLD layers were deposited at 800 °C at 0.3 mbar of $pO_2$ with a pulse frequency of 5 Hz. The PLD-targets were fabricated by Oxolutia SL (Spain) and consist of pressed and sintered, stoichiometric YBCO powder at 87% density. CSD and PLD films were either cooled in dry growth atmosphere (as-deposited) or directly oxygenated during cooling by increasing the $pO_2$ in the reactor chamber, and, in the case of CSD, adding an additional dwell between 450 and 550 °C.

Metallic counter electrodes and top current collector layers were paint brushed using Ag and Au paste, respectively, diluted with isopropanol. Porosity and good electrical conductivity was induced by a short drying and sintering process at 300 °C in lab air.

The highly epitaxial growth was confirmed and the *c*-axis lattice parameter was determined using a Bruker D8 Advance series II diffractometer (monochromatic Cu-K$_{\alpha 1}$ radiation, λ = 1.541 Å). The surface was studied using an FEG ZEISS GeminiSEM 300 microscope in secondary and backscattered electron mode with an accelerating voltage of 3 kV in high vacuum.

Electrochemical treatments were performed at LMGP using two different high temperature setups, including a Nextron temperature cell, equipped with a 1/2" ceramic heating stage and six electrical probes, and a home-built electrical probe setup placed within a tubular furnace. Electrochemical measurements were performed using a Keithley 2400 sourcemeter and a Biologic SP-150 potentiostat. *In situ* XRD measurements were performed on a RIGAKU Smartlab diffractometer in BB geometry with a DTex 1D detector, two Ni-K$_\beta$ filters at the source and an Anton Paar DHS 1100 heating stage, closed with a graphite dome. A home-built extension allows to insert cables for electrical measurements.[74] For combined electrochemical treatments and *in situ* conductivity measurements, metallic Ti(5 nm)/Au (100 nm) top electrodes were fabricated in Nanofab clean room facilities using photolithography and evaporation.

Low temperature electrical and superconducting properties were analysed at ICMAB and Institut Néel. The low temperature resistivity, $\rho$, and charge carrier density (via the Hall effect) were measured in a Physical Property Measurement System (PPMS, Quantum Design) using the Van der Pauw configuration. Electrical measurements were averaged over two permutations of the electrical contacts and positive and negative excitation currents in DC mode (between 0.1 and 5 mA). The critical current density was obtained via SQUID magnetometry (Quantum Design), as explained elsewhere.[16]

# Acknowledgments

This research was funded in part by the Austrian Science Fund (FWF) 10.55776/PAT2412325. This work received government funding managed by the French National Research Agency under France 2030, reference number "ANR-24-EXSF-0004". The authors acknowledge M. Eisterer (TU Wien) for careful revision of the manuscript. A.S. would further like to thank M. Eisterer for his encouragement to pursue on this topic. Authors acknowledge NANOFAB facilities of the Néel Institut as well as technical assistance from the CMTC platform.

# Data Availability Statement

The data that support the findings of this study will be made available in zenodo at https://doi.org/10.5281/zenodo.17607602

# Declaration of Competing Interest

The authors declare no known conflict of interest.

# Author Contributions

A.S. developed the original idea of this manuscript and designed the methodology and experimental studies. A.K. and C.P. synthesized the samples. A.S., H.R. and A.K. performed the experimental work. A.S. analysed the data and prepared the manuscript. A.B., A.S. and T.P. acquired the funding for research and personnel. M.B., T.P. and X.O. contributed to the discussion of the results and all authors contributed to the revision of the manuscript.

# Supplementary Information:

# Fast track to the overdoped regime of superconducting YBa$_2$Cu$_3$O$_{7-\delta}$ thin films via electrochemical oxidation

Alexander Stangl[1,2,3*], Aiswarya Kethamkuzhi[4], Hervé Roussel[3], Cornelia Pop[4], Xavier Obradors[4], Teresa Puig[4], Mónica Burriel[3] and Arnaud Badel[5]

* alexander.stangl@grenoble-inp.fr

[1] Université Grenoble Alpes, CNRS, Grenoble INP, Institut Néel, 38000 Grenoble, France

[2] TU Wien, Atominstitut, Stadionallee 2, 1020 Vienna, Austria

[3] Université Grenoble Alpes, CNRS, Grenoble INP, LMGP, 38000 Grenoble, France

[4] Institut de Ciència de Materials de Barcelona (ICMAB-CSIC), 08193 Bellaterra, Barcelona, Spain

[5] Université Grenoble Alpes, CNRS, Grenoble INP, G2ELab – Institut Néel, 38000 Grenoble, France

**Supplementary Information - Note 1: Overdoping**

In the field of high-$T_c$ superconductivity and in particular for the triple-perovskite YBCO, the term doping is used somewhat ambiguously. Electronic doping, $p$, refers to the number of charges (*i.e.* holes) per Cu ion in the CuO$_2$-planes. The critical temperature, $T_c$, forms the well-established superconducting dome in the $T - p$ phase diagram of cuprates, see main text Figure 1(a), which is commonly divided into three regions. In the underdoped regime, the critical temperature increases with $p$ until it peaks at optimal doping with $p = 0.16$ holes/Cu. In the overdoped regime, $T_c$ decreases again, while the superconducting condensation energy (and thus the critical current) keeps increasing.

Oxygen doping on the other hand refers to the modulation of the flexible oxygen content in the CuO-chains. In a simple picture, each incorporated oxygen ion (O$^{2-}$) is linked to the release of two free electron holes (h$^+$). Via a non-trivial charge exchange mechanism,[75] some of these holes are transferred from the CuO-chains to the CuO$_2$-planes, leading to an increase of $p$ with the oxygen content. As a matter of fact, neither the electronic doping value nor the absolute oxygen content are easily accessible (especially in YBCO thin films) and thus are frequently approximated via $T_c$, the *c*-axis lattice parameter, the thermopower or the charger carrier density, $n_H$, with each quantity associated with specific limitations and uncertainties. Here, we utilized the charge carrier density, as it was recently demonstrated to be an (injective) function of the electronic doping, $n_H = f(p)$,[76] with optimal doping corresponding to about $n_H(T = 300\text{K}) \approx 8 \cdot 10^{21}$ cm$^{-3}$.[16,66,67] Consequently, higher charge carrier densities can be associated with the overdoped state.

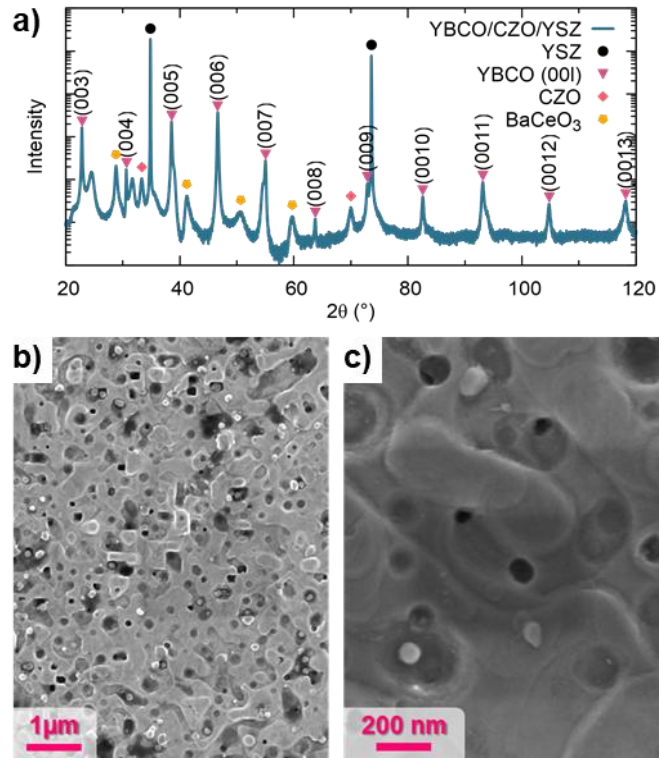

SI-Figure 1: (a) XRD diffraction pattern confirming highly epitaxial growth of YBCO (CSD) on YSZ substrate using $(Ce,Zr)O_{2-\delta}$ buffer layer. Some reactivity between Ba and the CZO buffer is observed via the formation of $BaCeO_3$. (b & c) Secondary electron SEM images of YBCO (CSD) surface.

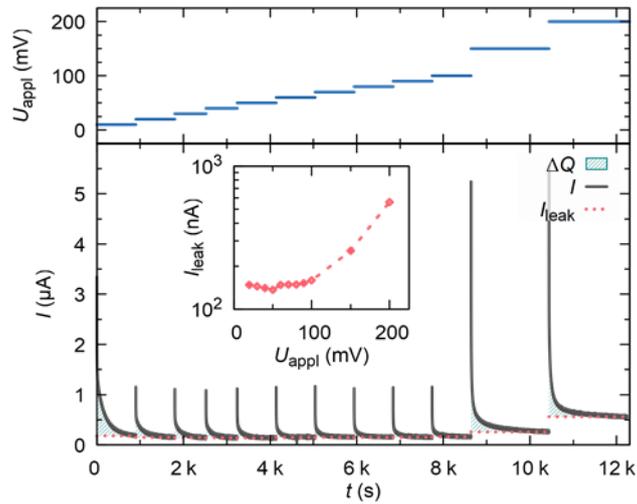

SI-Figure 2: Electrochemical titration of CSD grown YBCO at 375 °C upon stepwise increase of the applied voltage (top panel). The inset shows the leakage current as function of the applied voltage.

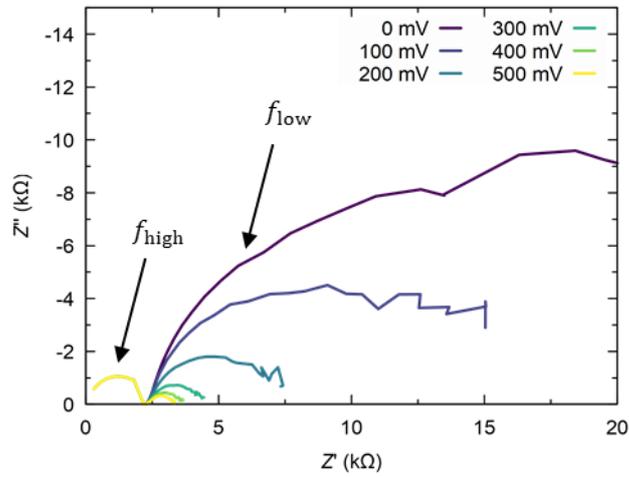

SI-Figure 3: Electrochemical impedance spectroscopy (EIS) study of YBCO/CZO/YSZ with a porous Ag counter electrode at 400 °C in 1 bar of $O_2$ with a 50 mV AC signal and various DC polarizations. Two main contributions can be distinguished: the semi circles at high frequencies, $f_{high}$, is ascribed to the YSZ electrolyte, the low frequency semi-circle corresponds to YBCO. As expected, the YSZ contribution remains invariant upon increasing DC biases, while the YBCO polarization resistance strongly decreases. This limits the accessible overpotential range. The contribution of the porous Ag counter electrode is expected to be much smaller and therefore not distinguishable in this preliminary data.

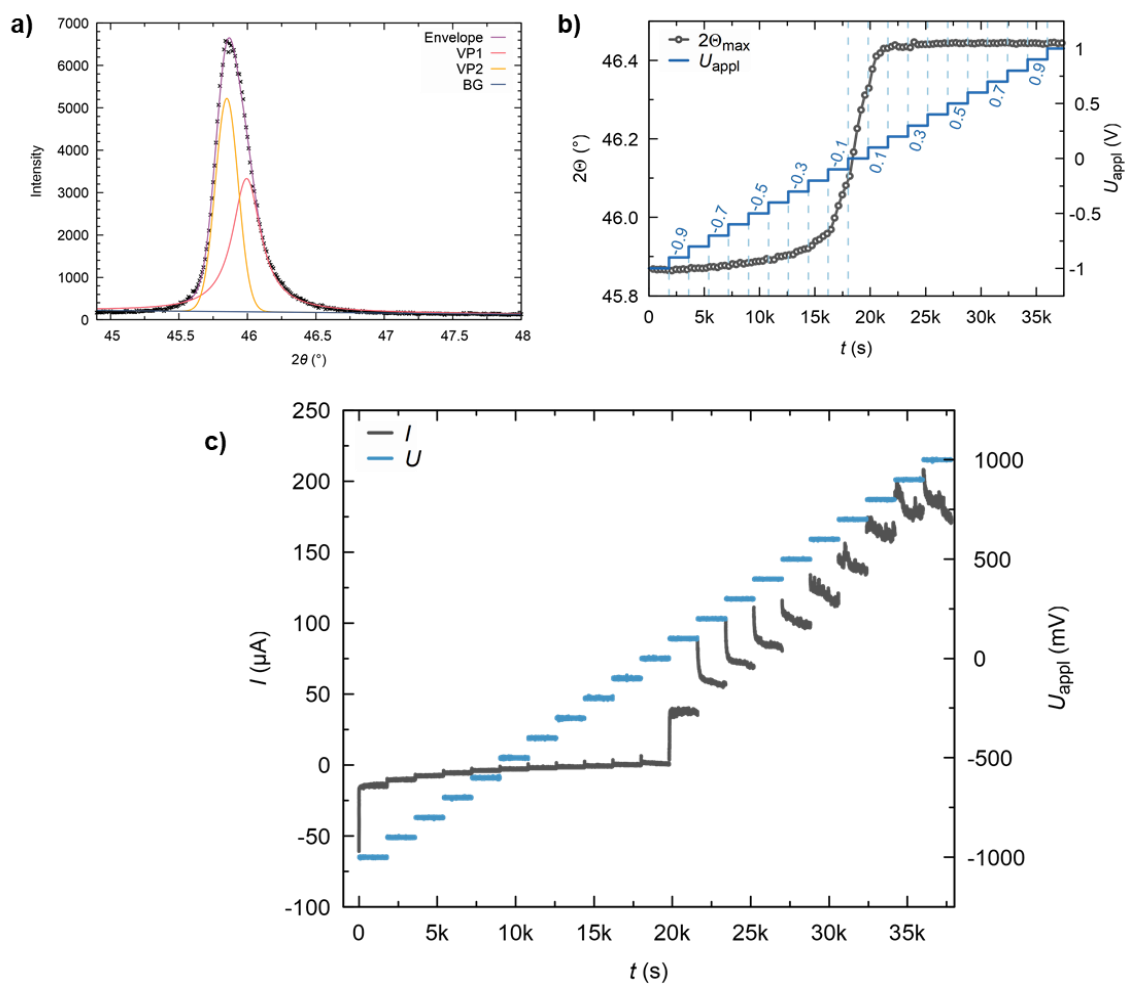

SI-Figure 4: *In situ* XRD measurements at 380 °C. (a) The process of XRD peak fitting: to account for the observed asymmetry, the maximum position is obtained from the envelope of two Voigt profiles. (b) Time evolution of the (006) reflection. Changes in 2Θ are caused by a stepwise increase of the applied voltage, shown in blue, and the resulting incorporation of oxygen. The corresponding titration curve is shown in (c).

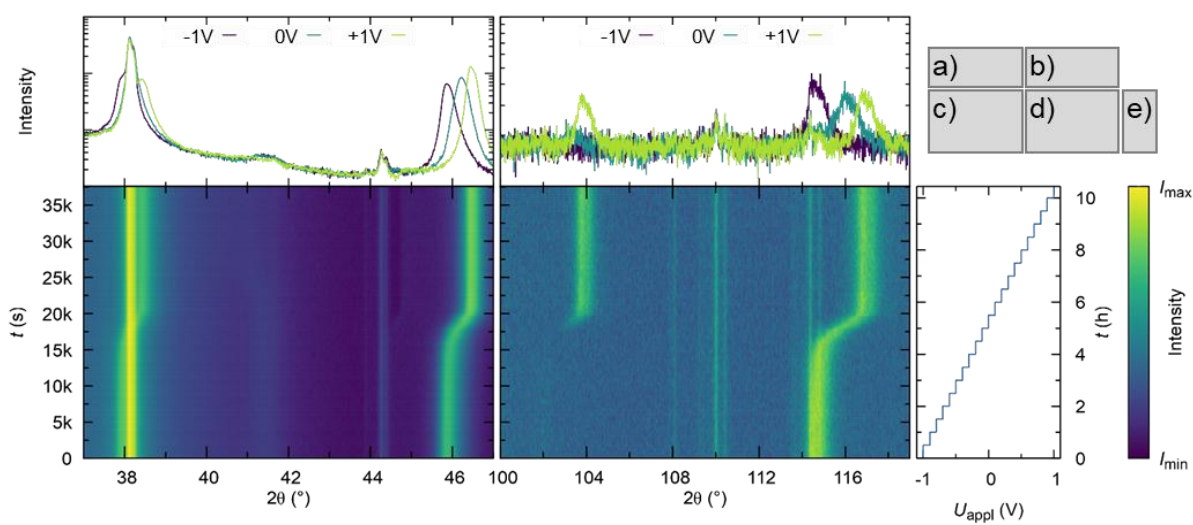

SI-Figure 5: Full set of collected *in situ* XRD data at 380°C: (a & b) End and mid point diffraction pattern at -1, 0 and +1V. (c & d) Time resolved intensity maps for the for the recorded 2Θ ranges and the corresponding $U(t)$ profile in (e).

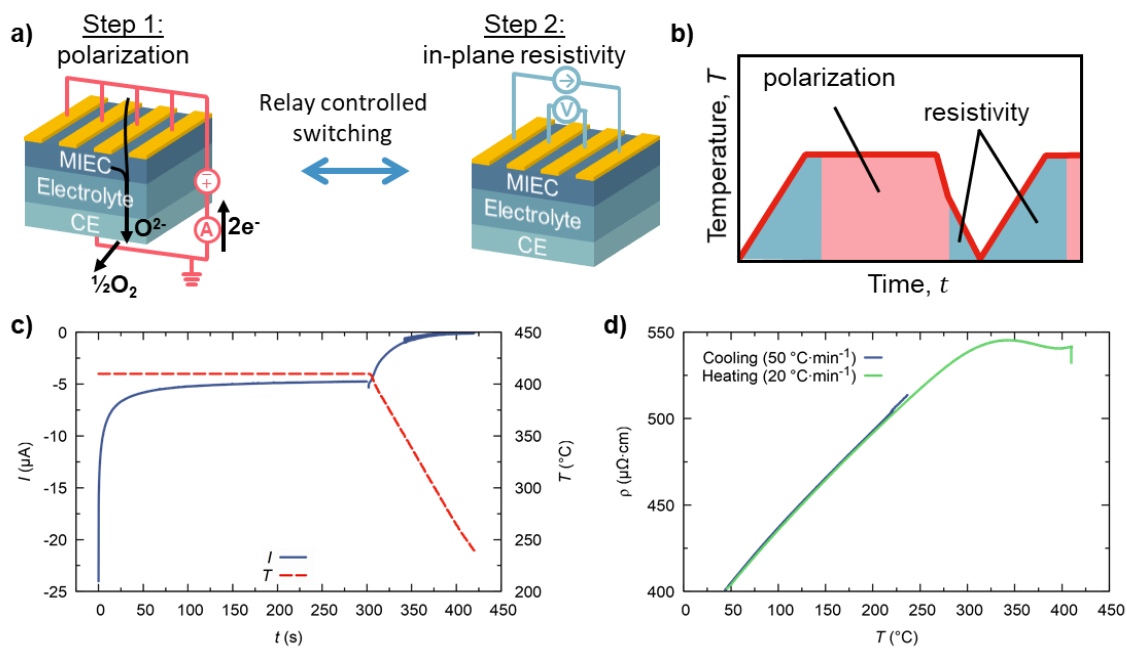

SI-Figure 6: (a) Advanced electrochemical cell for combined electrochemical and electrical analysis. Specifically designed top electrodes allow to switch between two configurations: in a first step, the MIEC layer is polarized out-of-plane by short circuiting the four metallic top electrodes and applying a voltage against the bottom counter electrode to pump oxygen out of the MIEC layer. In a second step, the electrical out-of-plane connection is cut and the four top electrodes are contacted individually to allow for in-plane resistivity measurements. Switching is performed using a digital relay. (b) Temperature profile illustrating the repeated, sequential measurement steps. (c) Out-of-plane titration process at 410 °C with an applied voltage of -100 mV. The current first saturates under isothermal conditions and then rapidly vanishes during the fast cooling above 250 °C. Zero current indicates that no more oxygen is moved through the electrolyte and the oxygen stoichiometry of the MIEC is frozen in. (d) Below 250 °C, the electrode configuration is switched to measure the in-plane resistivity as function of the temperature. The overlap of cooling and heating curves confirms that the MIEC oxygen content is not changed at these low temperatures. The deviation from linearity during heating above 300 °C marks the onset temperature of oxygen exchange reactions of the native YBCO surface.

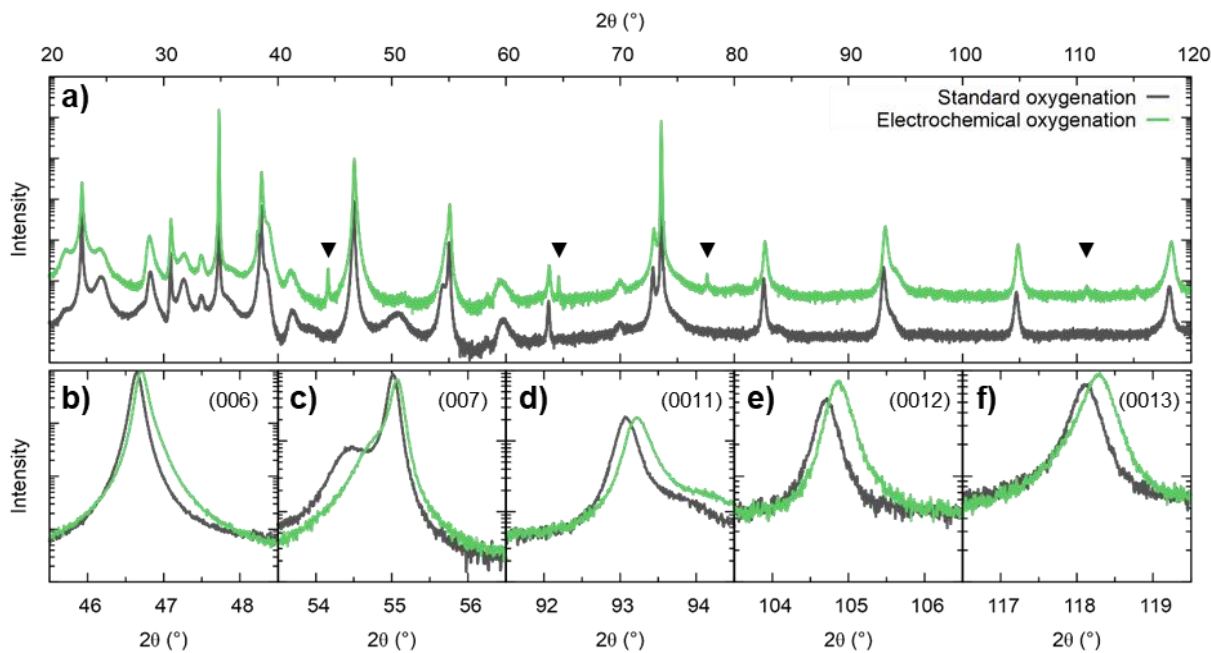

SI-Figure 7: Room temperature XRD analysis of CSD grown YBCO film prior and post electrochemical treatment. The sample was initially approximately optimally doped using the standard oxygenation process. Subpanels (b-f) show magnified regions around the (00l) reflections. Black triangles in (a) mark peaks arising from metallic current collector.

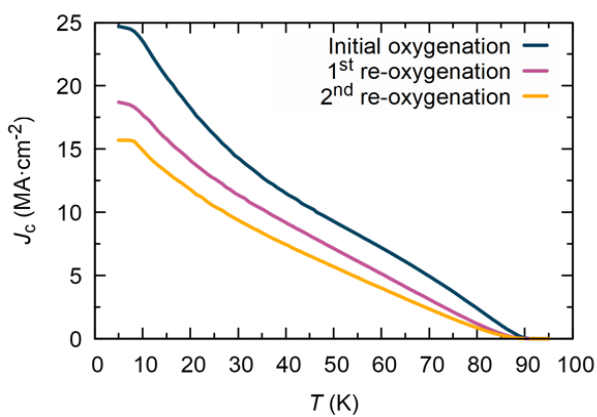

SI-Figure 8: Evolution of the critical current density of a YBCO film (CSD) upon repeated exposure to standard oxygenation conditions at 450 °C in flowing $O_2$. After each heat cycle the $J_c(T)$ decreases in magnitude over the full temperature range.

| Sample | 1 | 2 | 3 | 4 | 5 | 6 | 7 |
|---|---|---|---|---|---|---|---|
| Technique | CSD | CSD | CSD | PLD | PLD | PLD | PLD |
| Oxygenated prior to EC treatment | Yes | Yes | Yes | Yes | Yes | No | No |
| Initial $J_c$ | 32.5 | 29.3 | 15 | 54.5 | 52.8 | 10.9 | 10.4 |
| Final $J_c$ | 13.2 | 15.8 | 3.6 | 39.9 | 44.9 | 54.0 | 48.5 |

SI-Table 1: Critical current densities (in MA·cm² at 5 K) prior and after electrochemical treatments for CSD and PLD grown samples.